\documentclass[prb,preprint]{revtex4-1}

\usepackage{amsmath}
\usepackage{amsfonts}
\usepackage{graphicx}

\begin{document}

\title{Action Physics}

\author{Lachlan P. McGinness}
\author{C. M. Savage}
\email{craig.savage@anu.edu.au}
\affiliation{Physics Education Centre, 
Research School of Physics and Engineering,
The Australian National University,
Canberra ACT 2601, Australia}

\date{\today}

\begin{abstract}
More than a decade ago Edwin Taylor issued a ``call to action" that presented the case for basing introductory university mechanics teaching around the principle of stationary action. \cite{Taylor2003} We report on our response to that call in the form of an investigation of the teaching and learning of the stationary action formulation of physics in a first-year university course.  Our action physics instruction proceeded from the many-paths approach to quantum physics through to ray optics, classical mechanics, and relativity. Despite the challenges presented by action physics, students reported it to be accessible, interesting, motivational and valuable. 
\end{abstract}

\maketitle

\section{Introduction}
\label{Introduction}
In a 2003 guest editorial in this journal Edwin Taylor argued for unifying quantum, classical and relativistic physics instruction around the principle of stationary action. \cite{Taylor2003} We call this unified approach ``action physics''. In the following we describe our implementation of a version of it in first-year university teaching.

Action physics arises from the fact that the fundamental characterization of a system is in terms of energy. Taylor's call sought to bring into the classroom the evolution of fundamental physics away from the concept of force and towards the concept of energy. \cite{Jammer,Lanczos}  This evolution is fully realized in classical analytical mechanics and in quantum mechanics where a system is specified by its total energy or Hamiltonian. Taylor argued that the concept of energy is simpler and more useful than that of force. Indeed, the scalar energy is mathematically simpler than the vector forces, which is why analysing complex mechanical systems is often easier using analytical mechanics than it is using Newtonian mechanics. Furthermore, outside physics classrooms, energy is an increasingly important concept in human affairs.

At the Australian National University (ANU), we also desired to increase student engagement by presenting them with physics they had not previously studied and with new ways of understanding the world. Furthermore, cognitive science tells us that ``chunking'' of knowledge into broad categories both improves learning and is characteristic of expert behavior. \cite{Reif,Docktor} Action physics achieves this by uniting different areas of physics under the stationary action principle, and explaining classical physics in terms of quantum physics.

In this paper we describe our work investigating the teaching and learning of introductory action physics, which has been included in our first-year physics course for a number of years, broadly following Taylor's model. This is an ambitious program, as experience warns that action physics might be too sophisticated for first-year students, and that its proper place is in later years when students are better able to appreciate its beauty and subtlety. In section \ref{Student Attitudes} we provide evidence from a student survey that this was not so for our students. 

Action physics may increase the mathematical and conceptual demands on students. 
Taylor \cite{TaylorCIP} and others \cite{Resources} have argued that computer visualization can address some of these challenges. We have found that mathematical packages, such as Mathematica, \cite{Mathematica} capable of symbolic manipulation, numerical solutions and graphing, can indeed enhance students' capabilities.

A body of literature about implementing action physics instruction has been available for many years, \cite{TaylorCIP, Hanc2003, Ogborn, Neuenschwander} including in the April 2004 theme issue of this journal, on classical mechanics. \cite{Moore, HancAJP2004a, HancAJP2004b, HancAJP2004c} This literature was the foundation on which we built our instruction.
Resources to support instruction are also available, \cite{Resources, QEDvideo} including assignments and computational laboratories in the supplemental material. \cite{supplement}  To help manage the implementation of action physics we took a physics education research approach. \cite{McDermott, ethics} In particular, we developed a concept inventory for action physics. \cite{McGinness}

Action physics is not common in first-year university teaching. Indeed, the lack of broad-based innovation in physics content for physics majors is striking, particularly considering the widespread adoption of innovative pedagogy, such as active teaching methods. \cite{Active} Perhaps this is in part because the restructuring required to teach action physics is challenging. Content must be removed to accomodate it, without compromising later year courses. On the positive side, Moore has discussed the opportunities for subsequent courses created by an action physics foundation. \cite{Moore} 

One of the challenges of introductory instruction is selecting content from a wide range of fundamental and applied physics, not all of which can be done justice in the first year of instruction. We may wish to teach relativity and quantum mechanics because they are amongst the hard won jewels of human culture, and thermodynamics and fluids because they are so useful. The choices described in this paper are not currently part of mainstream physics education, but we have shown they can work. We hope and expect that our work may inform instructors with a variety of experiences and preferences.

In the next section we describe the educational context of our course and its special characteristics. In section \ref{Action Physics} we define more precisely what we mean by action physics. Section \ref{Student Attitudes} reports the results of a survey of students' attitudes to action physics. We conclude with a discussion of achievements and remaining challenges.

\section{The Course}
\label{The Course}

Action physics has been incorporated into Physics 2, the second course required for physics majors at the ANU. \cite{phys1201}  The students are high-achievers at the national level, and the ANU physics program aims to provide a high level of instruction. \cite{PECMission} We have not fully implemented Taylor's vision of bypassing Newtonian physics since almost all students have taken Physics 1 in the previous semester, which consolidates their pre-tertiary instruction in Newtonian mechanics, electrostatics and magnetostatics. \cite{phys1101}  Newtonian mechanics, for which force is a fundamental concept, provides an understanding of the concepts of kinetic and potential energy that are essential for formulating action physics. Physics 1 is taken by most engineering students but Physics 2 by only a few. Each course is a quarter of a full-time load for a semester. The students are required to take mathematics courses covering: complex numbers, differential equations, and partial differentiation.

Action physics was first taught to Physics 2 in 2006 and 2007 to classes of about 70 students, after which it was displaced by other content. In 2013 it was taught again, but as a course option for extra credit to a group of 27 out of the 115 students. This allowed close observation of student learning. Informal evaluations from these students and from academics were positive overall, with two students commenting on the ``beauty of the action principle'' in open ended survey questions. There was no evidence that the material was too sophisticated for the group. In particular, the level of mathematics was not identified as a problem, although the workload and time required to learn Mathematica were. The results we report are drawn only from our experience with this course and hence may not apply to courses requiring different levels of preparation in physics and mathematics. Nevertheless, our students are diverse, and hence much of our work may be adaptable to other teaching contexts.

After incorporating the lessons learnt from the 2013 class into our teaching plan, action physics was included in the 2014 course for all 110 students. The action physics instruction was a module of eight 50 minute lectures, a 90 minute computational laboratory, and two 50 minute tutorials. The lectures included class-response questions.  It was assessed by two homeworks, a lab logbook and a final exam question. The progression of topics was: many-paths quantum physics, Fermat's principle, the stationary action principle for mechanics, Euler-Lagrange equations, general coordinates, symmetry, and relativity.

The students' learning of action physics was closely scrutinised, with some homework tasks designed to give us insight into their conceptual understanding.  The other topics in Physics 2 were: special relativity, electromagnetism, and waves and optics. Performance in the action exam question was comparable to that in the exam as a whole.

A concept inventory was developed during 2014, which involved working with academics at ANU and around the world to determine the key concepts of action physics. \cite{McGinness} The inventory development included observation of student thinking using think aloud interviews. \cite{Adams} The evaluation based on this information, as well as on standard ANU course surveys, was sufficiently encouraging that we taught action physics again in 2015. We also had  particularly positive feedback from a lecturer teaching a second-year physics class that included most of the group of 27 students who took action physics in 2013.

Presenting sufficient quantum physics to provide a foundation for classical physics within the constraints of the action physics teaching module proved the most challenging part of the pedagogy. Hence, in 2015 the computational laboratory was increased from 90 to 180 minutes, half of which dealt with many-paths quantum physics. Our limited goal was to provide just the necessary foundational conceptual understanding, but not the capacity to actually calculate quantities, such as probabilities. Those students who wished to know more were referred to Feynman's popular book \textit{QED}.  \cite{QED}

\section{Action Physics}
\label{Action Physics}

In this section we describe what was taught in the action physics module. Background material and further information may be found in the action physics education literature. \cite{TaylorCIP, Hanc2003, Moore, HancAJP2004a, HancAJP2004b, Ogborn, Neuenschwander} As the rationale for action physics includes presenting a unified view of physics, we started our instruction with the foundation, quantum physics. Specifically with the many-paths formulation of quantum physics, introduced by Feynman in 1948. \cite{FeynmanRMP, FeynmanHibbs} It may seem ambitious to introduce quantum mechanics with a formulation that is often only taught to advanced students. The case for doing so is made in Feynman's book \textit{QED}, including in Zee's introduction to the 2006 edition. \cite{QED} In brief, it is that the many-paths formulation is both more fundamental and conceptually simpler than that based on wavefunctions. However, it is technically much more difficult when it comes to calculating things. Hence, it has advantages when only the conceptual basis of quantum physics is taught. Just as importantly, it has a straightforward connection with classical dynamics. \cite{TaylorCIP, Ogborn, Dirac} The optics version was introduced in his popular book \textit{QED},  \cite{QED} and developed further by others. \cite{deGrooth,Field}

Our first technical quantum physics course is in the students' second year and takes a conventional approach. \cite{Griffiths} However, many-paths quantum physics is developed technically in our largest third-year physics course, which about half of Physics 2 students go on to take.

Quantum physics describes the world by the probability for a system to make a transition from state A to state B, $\textrm{Pr} ( \textrm{A} \rightarrow \textrm{B} )$. This is expressed in terms of the transition amplitude $\alpha ( \textrm{A} \rightarrow \textrm{B} )$ by the square of its modulus: $\textrm{Pr} ( \textrm{A} \rightarrow \textrm{B} ) = | \alpha ( \textrm{A} \rightarrow \textrm{B} ) |^2$. In the many-paths formulation amplitudes are calculated by a sum over all possible paths from A to B: 
\begin{equation} \label{amplitude}
\alpha ( \textrm{A} \rightarrow \textrm{B} ) = N \sum_\textrm{all paths} \exp ( i S_\textrm{path} / \hbar ) ,
\end{equation}
where $N$ is a normalization factor. $S_\textrm{path}$ is the action calculated for each path, that is for each possible way for the system to get from A to B, and $\hbar$ is the reduced Planck constant. 

From a naive perspective, such as the students', this appears to be an arbitrary and abstract theoretical construct. Chabay and Sherwood have argued that something similar is true of first-year electromagnetism. \cite{ChabayEM} Many students find the concepts and mathematics underpinning the integral formulation of Maxwell's equations abstract and sophisticated. The arguments for teaching electromagnetism at this level also apply to action physics, which is similarly abstract and mathematical.

\begin{figure}[]
\includegraphics[width=14cm]{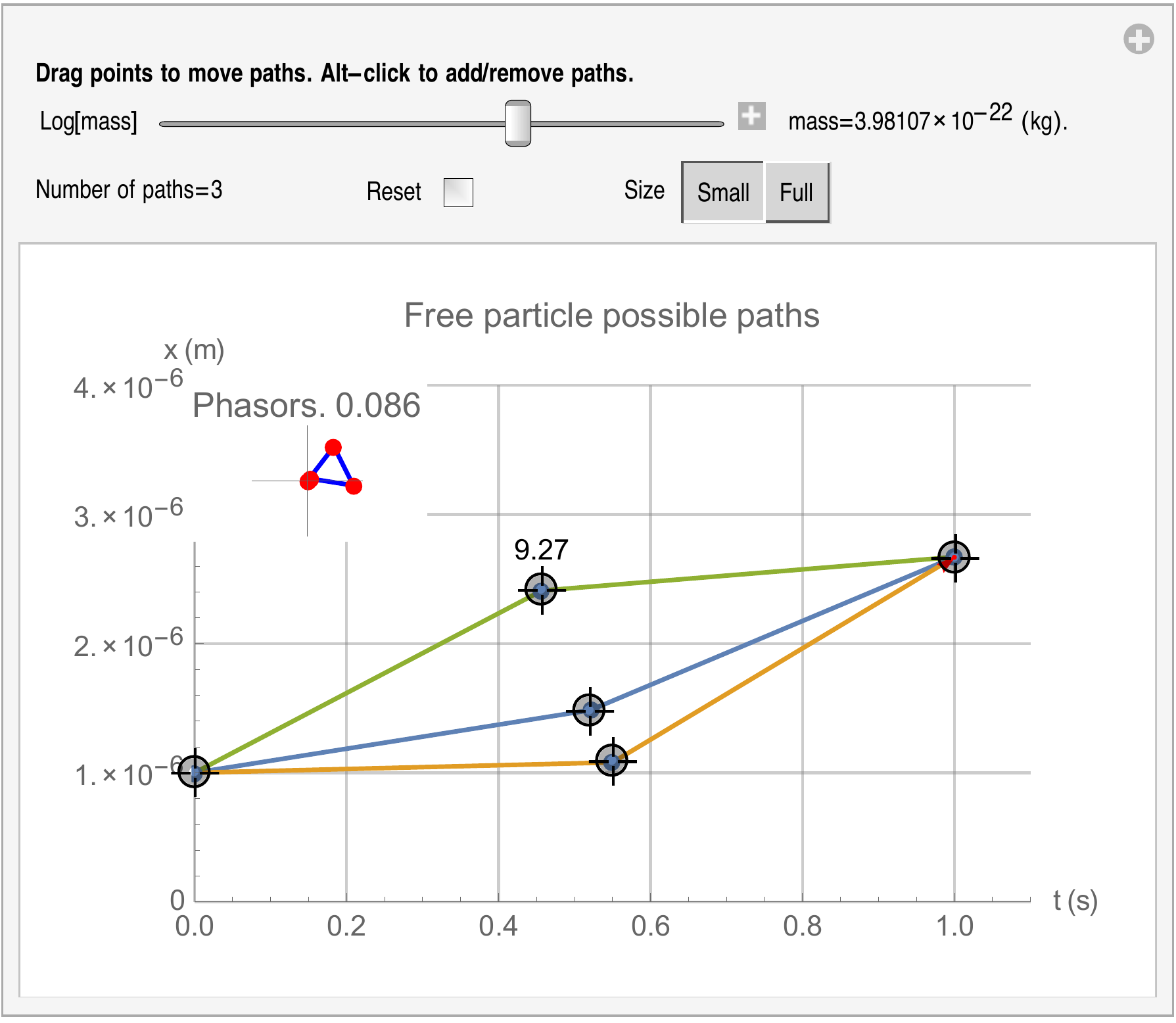}
\caption{Screenshot from the quantum many-paths Mathematica notebook used in the computational laboratory, available in the supplemental material. \cite{supplement} The notebook includes a complete description. Briefly, it is a space-time diagram with position on the vertical axis and time on the horizontal axis that shows three free-particle, piecewise-linear paths. Each path is defined by the fixed starting and ending events and an intermediate event that may be moved interactively. These particular paths destructively interfere to give almost zero amplitude. This is shown in the upper left inset that represents the complex amplitude of each path as a phasor; added nose-to-tail. The notebook is based on the work of Taylor \textit{et al.}, \cite{TaylorCIP} which should be consulted for further explanation. }
\label{destructive interference}
\end{figure}

A critical part of the development is showing how classical physics arises from the quantum mechanical transition probabilities when the action is large compared to $\hbar$ for all paths, $S_\textrm{path} / \hbar \gg 1$. Then the exponential in the amplitude summand, Eq.~(\ref{amplitude}), rotates rapidly around the unit circle in the complex plane as the path varies over the sum, see Fig.~\ref{destructive interference}. Consequently, the terms destructively interfere, except around paths for which the action is stationary, that is for which first order changes in the path produce no first order change in the action. \cite{TaylorCIP, Ogborn, QED} The connection with classical physics is made by identifying the stationary action path with the classical trajectory. In contrast, for quantum mechanical systems satisfying, $S_\textrm{path} / \hbar \approx 1$: destructive interference does not occur, all paths contribute to the amplitude's value, and classical physics is not a valid approximation.

Having ``derived'' classical physics from quantum physics, we considered three specific cases: light ray propagation, simple mechanical systems, and relativistic free particles. In the ray optics approximation the action is simple - the product of the photon energy $\hbar \omega$, where $\omega$ is the light's angular frequency, and the time for light to travel the path, $t_\textrm{path}$: $S_\textrm{light} = \hbar \omega t_\textrm{path}$. This is of course the action for light in the quantum mechanical sum over paths, Eq.(1). The initial and final states are spatial locations of the ray. Fermat's principle says that a classical ray is one whose travel time between points A and B is stationary.

For classical mechanics the action is defined in terms of the system's kinetic energy $K$ and potential energy $V$ by the time integral:
\begin{equation} \label{action}
S_\textrm{mechanics} = \int_{t_A}^{t_B} K-V \, dt .
\end{equation}
The initial and final states are events, in the relativistic sense, defined by their space and time coordinates: $(x_A,t_A)$ for event A and $(x_B,t_B)$ for the later event B.

Euler-Lagrange equations for the stationary path $x_s(t)$ were derived from the stationary action principle using ordinary calculus by considering the variation of the action with respect to the infinitesimal variable $\epsilon$ for the paths $x(t) = x_s(t)+\epsilon f(t)$, where $f(t)$ is any function that is zero at times $t_A$ and $t_B$. \cite{HancAJP2004b} 

One of the assessment problems required finding the Lagrangian, and hence the Euler-Lagrange equations, for the double pendulum. A Mathematica notebook was provided to assist with the calculations. \cite{Mathematica, supplement} This helped keep the students' focus on the physics rather than the mathematics.

Relativistic free particle kinematics is based on the principle of maximal aging: \cite{TaylorandWheeler} the physical path of a free particle in special and general relativity is that which maximises the proper time $\tau_\textrm{AB}$ between the starting and ending events A and B,
\begin{equation} \label{maximal aging}
\tau_\textrm{AB} = \int_\textrm{A}^\textrm{B}  \, d\tau ,
\end{equation}
where $\tau$ is the particle's proper time, and the integral is along a particular world-line. \cite{GR} Multiplication by the particle's rest mass energy $m c^2$ gives the integrand the units of energy, as required for an action principle:
\begin{equation} \label{maximal aging rest mass}
S_\textrm{relativistic} = m c^2 \tau_\textrm{AB} = \int_\textrm{A}^\textrm{B} m c^2  \, d\tau .
\end{equation}
For this to be useful, special relativity must be taught separately, including the relationship between proper time and coordinate time, $t$. Then the principle of maximal aging is helpful for understanding time dilation and the twin paradox. \cite{Taylor2003,TaylorandWheeler}

\section{Student Attitudes}
\label{Student Attitudes}

At the end of the 2014 action physics module 93 students completed the attitude survey included in the supplemental material. \cite{supplement} Its primary purpose was to investigate how students perceived the difficulty and value of action physics compared to that of the other physics they had studied. 

The first question was: ``Do you think action physics should be included in Physics 2 next year?''. 89 students (96\%) chose the answer yes and 4 (4\%) chose no. Another question was: ``In relation to other parts of Physics 1 and Physics 2, how has studying action improved your understanding of physics?''. Responses were on a seven-point Likert scale from: 1, ``not at all''; to 7, ``much more than other topics''; with a mid-point of 4, ``about the same''. There were 8 (9\%) negative responses ($<4$), 20 (22\%) neutral responses (4), and 65 (69\%) positive responses ($>4$). This indicates that most students believed they understood physics more through studying action than other topics.

Students were also asked to select from a list as many responses as applied to the question: ``What aspects of action physics did you enjoy learning the most?''. Of the six options the two most commonly chosen were: `` A new perspective on physics'' (84 or 90\%), and ``Seeing the link between different areas of physics'' (59 or 63\%). This suggests that we achieved our goals of a unified approach to physics and of presenting new ways of understanding the world. 

In response to the question, ``What did you find most challenging about learning action physics?'', out of five non-exclusive options the most commonly chosen was ``Using Mathematica'' (58 or 62\%). The next two most common were ``Calculus of variations and other mathematics'' (40 or 43\%) and ``Quantum mechanical concepts'' (39 or 42\%). The challenge of mastering Mathematica software relates more to our physics major program than specifically to the action physics module. \cite{Mathematica}

Students were also asked to rate different topics on a seven-point Likert scale, in the dimensions of relative difficulty, interest and motivation to learn. The results are shown in Table I. Action physics was rated between relativity and electromagnetism in all dimensions. This is evidence that action physics fits comfortably in the first-year curriculum, being regarded by students similarly to special relativity.

\begin{table}[t]
\centering
\caption{Student attitudes of the 2014 Physics 2 class towards topics covered in Physics 1, and in Physics 2 by the end of the action physics module. Class means over a 7-point Likert scale: difficulty (1-easy, 7-difficult); interest and motivation (1-low, 7-high). The neutral benchmark (4) was Newtonian mechanics in Physics 1. The statistical errors in these means are between 0.1 and 0.16.}
\begin{ruledtabular}
\begin{tabular}{l c c c }
Topic & Difficulty & Interest & Motivation \\
\hline	
Relativity & 4.6 & 6.0 & 5.3  \\
Action & 5.4 & 5.9 & 5.1 \\
Electromagnetism & 5.9 & 3.9 & 3.7 \\
\end{tabular}
\end{ruledtabular}
\label{bosons}
\end{table}

One of the open ended comments in the survey gives a powerful student perspective on including action physics in first-year courses:
\begin{quote}
``I want to thank you for the opportunity to learn something so profound and spectacular. When I was 14 I read the Elegant Universe by Brian Greene - everything he said was conceptually incredible, but he quite deliberately gave non-mathematical explanations for every topic. Great for a young and/or unexposed mind, but it left quite a few open-ended questions - many of which you have provided something of an answer for. Thank you for deepening my understanding of the world in which we live, even by a small amount. I'm inspired.'' 
\end{quote}

\section{Conclusion}
\label{Conclusion}

The evidence presented shows that action physics has been successfully taught in a particular first-year university course. The success is attested to by both teaching academics and students, and encompasses content and motivation. It was facilitated by the use of software tools to address conceptual and mathematical difficulties. 

Reports of other cases where teaching action physics earlier than is customary has been attempted are almost absent from the literature \cite{Taylor2003}. However, Ogborn has reported that many-paths quantum physics was a successful part of a widely used UK ``Advancing Physics'' high school A-level course around the turn of the millennium. \cite{OgbornGirep2006} The preparation of A-level students at that time may well have been similar to that of our Physics 2 students.

An important motivation for our teaching of action physics was to ensure that our content reflected current understanding and practice in physics. As quantum physics is a well established foundation we developed ways of conveying its conceptual and mathematical structure at the introductory level. This led naturally into the stationary action formulation of classical physics. We found that the sophistication this demands of students is comparable to that required by conventional calculus based first-year electromagnetism.

However, challenges remain. Firstly, the electromagnetism part of the curriculum is outside the domain of action physics. Secondly, it does not naturally incorporate friction, a central part of everyday experience.  It is a theory of fundamental interactions, and incorporating friction requires the atomic theory of matter. In practice, it must be added as a phenomenological term to the dynamical equations. \cite{Goldstein}

Over a number of years we have approached action physics instruction as a physics education research project, \cite{McDermott} and one of us (L.M.) completed a fourth-year thesis based on investigation of the 2014 class. Each year we have made evidence based decisions on how, and whether, to teach action physics. We have reached the point where its inclusion is a matter of competition with other topics and not of its value or effectiveness. 

We have demonstrated by example that a version of Taylor's vision of introductory action physics instruction can be made to work. Moore has described how later year courses might build on this. \cite{Moore} Although the particular characteristics of our course may be unusual, it serves a diverse spectrum of national and international students, and perhaps our experience might embolden others to take their introductory content in new directions.

\begin{acknowledgments}

We would like to thank the staff and students at ANU who have contributed to the development of the Action Physics module, and the referees for their thoughtful and constructive comments.

\end{acknowledgments}


\begin{thebibliography}{99}
\bibitem{Mathematica} ANU physics students have free access to Mathematica, which is used as a routine tool in the physics program.
\bibitem{GR} Since in general relativity there may be multiple geodesics between two events, the proper time is maximal with respect to neighboring paths.
\end{thebibliography}
\end{document}